# ANALYSIS OF INFORMATION TECHNOLOGIES USED TO INSURE WORKING EFFICIENCY OF PERSONNEL

*V.Ya. Vilisov, D.A. Dyatlova*
*University of Technology, Russia, Moscow Region, Korolev*
*vvib@yandex.ru*

***Abstract.*** *The work is devoted to a modern state, methods and tools of monitoring, assessment and prediction of the indicators showing physical condition of a person and his/her capabilities to perform work duties. The work contains an analysis of existing gadgets and software that allow tracking physical condition of personnel at the working place. The analysis showing significant interconnections and factors that determine a necessary level of working capacity and productivity of personnel allows organizing Work & Rest Schedule of employees in an effective manner.*

**Introduction**

Apart from current programs devoted to the near and deep space exploration, space agencies of different countries also perform advance research with a longer time lag. The advance research projects include manned explorations of remote planets, including Mars. One of the problems that should be solved is the creation of systems and technologies that allow astronauts to maintain their physical condition and working capacity at necessary level.

Even the richest countries of the world cannot boast of a perfect health care system. No matter how heavily medicine is invested, there is always a lack of skilled personnel. In some cases, the doctors simply do not have time to help an injured person. Unfortunately, critical condition of a chronic disease cannot be predicted. However, if all medical measures are taken in time, the patient can survive.

In order to perform a constant monitoring of the patient's health condition, the modern medicine needs the devices that would take certain readings from the person's body on a real time basis. Should there occur any threatening changes, the device would alert the patient or his/her physician. Some countries have already been introducing such devices into medical sphere [3].

The objective of the work is to perform an analysis of existing gadgets, applications and technologies that help people to track their health state. At that we consider two aspects:

- purely medical, within which the physiological indicators of a person are assessed and certain measures are taken in case of an abnormality or a certain prognosis;

- professional, when the person is an employee, whose vital signs should comply to some standard requirements.

The authors also consider some methodological aspects of working with the data that, within the monitoring process, enters corresponding storage places as well as some objectives and tasks that can be solved at the obtained data field.

**Performance Review of Health Monitoring Pilot Projects**

Monitoring people with various diseases is a market segment that grows very quickly. The number of patients that have one or several chronic diseases and who require remote monitoring services constitutes 200 mln people only in Western Europe and the USA [1]. The researchers predict [2] that more than 60 mln of devices monitoring people's health will have functioned within the mobile network by 2020, with the market volume reaching 18 billion US Dollars. Remote monitoring devices targeted at the patients with an irregular heartbeat, diabetes and chronic lung diseases are going to be in a maximum demand.

Mobile telehealth (mHealth) is one of the most promising directions in the sphere of new medical technologies [4-12]. Health care industry is of substantial interest for the network operators because here they can provide their customers with additional digital medical services. Lately mobile operators have been increasing their presence in this sphere. Due to a better network capacity, new generation networks are significantly expanding the capability of transferring medical data via mobile network channels, thus securing their prompt delivery, integrity and confidentiality. Moreover, modern communication channels allow organizing videoconferencing between a patient and his/her doctor or an operator who performs psychological testing or who measures physiological indicators.

Development of new generation networks (LTE) and current smartphone expansion provide a reliable technological base for the development of such projects. In this case, mobile medicine provides patients with different medical devices that they can use independently thus controlling their health and transmitting the monitoring results to medical centers.

There are thousands of applications created for iOS and Android platforms that can solve various medical tasks. Various additional devices and sensors transform a usual smartphone into a medical device that can take readings with a due precision. iPhone can help you to take an electrocardiogram, measure your blood pressure, check your eyesight or assess the risk of possessing skin cancer. And it is necessary to note that all above mentioned is done with an acceptable level of accuracy. The obtained data can be sent online to a subject matter expert who will provide you with his opinion of the matter.

Today, owners of smartphones and other mobile devices can already install applications that control their physiological parameters. Thus, for example, there is an application that performs an analysis of the birthmarks the person may have. One only needs to take a photo of the birthmark. Such an application would help you to go to the doctor in time to prevent skin cancer.

There are special applications that help you to check your hearing or eyesight, monitor pulse or get yourself thoroughly acquainted with any medicine.

Some applications that can be downloaded by users are shown in Table 1.

**Table 1 - Health Monitoring Applications**

| App Description | Android | iOS |
|---|---|---|
| Heartbeat Measurement | • Heart Rate Monitor<br>• Instant Heart Rate<br>• Runtastic Heart Rate | • Runtastic Heart Rate<br>• Cardiio – Pulsimeter |
| Hearing Check | • Hearing Check | • Petralex Hearing Device<br>• Hearing Check<br>• Hearing Aid |
| Vision Check | • Vision Check (andrew.brusentsov)<br>• Vision Check (healthcare4mobile) | • HD Vision Check<br>• Nearsightedness Check |
| Information about Medicines | • Medicines and Their Analogues<br>• Reference Book of Medicines and Diseases<br>• MedBox – Reference Book of Medicines | • Free Medicines<br>• Medicines from A to Z |

Many science & research institutes and medical companies have actively started developing devices and special applications for smartphones and tablets.

Experts predict that the burden of medical institutions will be significantly decreased in the future due to a mass usage of medical gadgets. Such devices shall release medical staff from standard and scheduled examinations, allowing them to allocate their time for those patients who are acutely in need of medical help.

Currently we already have the devices that allow the cardiologist, in real time and in any place of the world, to see at his/her smartphone the electrocardiogram of the patient and to monitor the rhythm.

We already have the sensors that continuously measure the blood sugar level. Today they are implanted under the skin. Still, we are not going to need the implants in the future, because the only necessary thing will be to set the maximum blood sugar level (higher than 75 and lower than 200); then the situation shall be controlled due to regular sampling performed by the sensor that would measure the blood level on a continuous basis, thus providing substantial help to people who suffer from diabetes.

There are more and more various tools in the field of sensor devices. And in most cases the technology becomes non-invasive, i.e. the one that does not need to be implanted into the body.

Table 2 contains examples of the devices that are available today.

Today, the speed of modern technology development allows providing medical services that are specifically tailored for each person. The approach presents us with the following main possibilities:
- a prompt diagnosis and an earlier detection of a disease;
- choosing an appropriate treatment, including usage of safe and effective medicines for each person;
- a more effective therapy, treatment monitoring and a diagnosis setting.

**Table 2 - Health Monitoring Gadgets and Their Description**

| Device | Summary |
| --- | --- |
| Zephyr BioPatch™ Wireless Device | A wireless device used for the patient monitoring (electrocardiogram; real heartbeat and its trend; a bio-conductor that measures water retention, which is critical when monitoring cardiac insufficiency; checking temperature, breathing, oxygen level, body position and movement). |
| Digital Thermometer Patch | A device that continuously measures temperature of a human body because even light variations can demonstrate the vessel work, thus helping to detect cardiovascular diseases. |
| Fitbit Force | An elegant bracelet made from a very light plastic. Its soft, but determined vibration at your wrist will wake you up in the morning. During the day it will count your steps, mounted stairs, walked kilometers and spent calories, while at night it will remember how many hours you were asleep and how often you tossed and turned in your sleep (sleep quality). Statistics, dynamics analysis and all social network functions are available. |
| iHealth BP5 | A wireless blood pressure monitor. iHealth BP5 is put on the forearm like an armband without any additional modules. An iOS-device connected with iHealth BP5 via a free application becomes its monitor, showing the data of the current blood pressure together with the statistics in general. |
| Lumoback | A posture monitor that can control and inform. Its main objective is to send you an occasional reminder regarding correcting your posture, which is controlled continuously. |

**Tasks of Health Performance Review**

Apart from using the above mentioned devices for medical purposes, they can also be used to monitor physiological and psychological indicators of healthy people whose work can cause risk for the lives of other people and/or risk of significant man-made accidents. Such jobs include bus drivers, electric train operators, air pilots, nuclear plant operators, astronauts etc. The parameters of Work & Rest Schedule (WRS) are very important for these work categories. We know many cases when non-compliance with a due WRS led to major accidents caused by tiredness or other factors that decrease the working capacity of personnel.

Therefore, the tasks of monitoring and forecasting of the personnel health deem very important for the purpose of maintaining the significant indicators at levels not lower than the acceptable ones.

The American and European [3] science area dealing with technical applications is called *Prognostics & Health Management* (*PHM*). The area contains a significant amount of information on discovering, recognizing and forecasting certain defects, which allows, without waiting for the failures and malfunctions, to provide a necessary level of the systems' working capacity. The "technocratic" approach can also be used in the sphere of medical and psychological monitoring of personnel. We shall call this direction *Personnel PHM (PPHM)*.

It can solve the following tasks:
1. indicator monitoring;
2. maintaining indicators within a set range (stabilization);
3. assessment of potential (maximum) capacities of personnel:
4. maintenance of extreme capacities of personnel.

The 1st and 3rd group tasks are used for evaluation but they can also contain a testing impact upon the person, thus obtaining necessary assessments.

Apart from the assessment, the 2nd and 4th group tasks contain the elements that are aimed at the development of controlling actions that provide necessary indicator values.

As a rule, the expanded list of the 1st and 2nd group sub-tasks includes the following:
- measurement of health state indicators in real time, accumulating the statistics for large time intervals together with a regular self-testing;
- processing measurement data in real time using special algorithms;
- integration of measurement data that was obtained from various sensors and psychological tests, computation of significant correlations existing between the indicators;

- computation of predictors and biomarkers demonstrating health deterioration together with a corresponding change of the Work & Rest Schedule (diet, sleep etc.);
- discovering and tracking negative trends, providing a preliminary notice to the monitoring subject or his/her managers;
- monitoring result output via friendly interface and its provision to a user who does not possess medical knowledge and skills.

*Biomarkers* and *Predictors* are the important elements that influence the effectiveness of solving the sub-tasks in the list.

Biomarkers are the indicators that can be measured in an objective manner and which could be used as indicators of physiological and psychological biological processes or as pharmacological responses to some therapeutic interventions. For example, they can include values of weight, blood pressure, pulse frequency, blood analysis, ultrasonic imaging, magnetic resonance imaging etc.

A substance that was introduced or found in the body can serve as a biomarker, thus being indicative of a specific illness or a presence of foreign bodies (for example, presence of specific antibodies can be indicative of a specific infection).

Currently, thousands of biomarkers are used to assess the health state, with their number constantly increasing.

Depending on the objectives of the subject examination, the following main types of biomarkers are distinguished:
- preventive biomarkers are used to discover a predisposition to contraction of a disease;
- verificatory biomarkers confirm presence of a disease at the subclinical stage;
- explorative biomarkers are used to identify a certain disease;
- state biomarkers are used to define a stage of the disease;
- prognostic biomarkers are used to evaluate the prognosis of the disease development, its possible outcome and assessment of the treatment's effectiveness;
- pharmacodynamic biomarkers are used to discover a specific pharmacological response that is necessary for the dosage optimization of the medicine, for example.

Predictor is a biomarker that can predict a favourable or unfavourable outcome of the disease or of the treatment's effectiveness.

**Health Monitoring Data Processing and Presenting Methods**

The logic of using biomarkers in a static condition (during separate and not interconnected time periods) is provided at Picture 1:

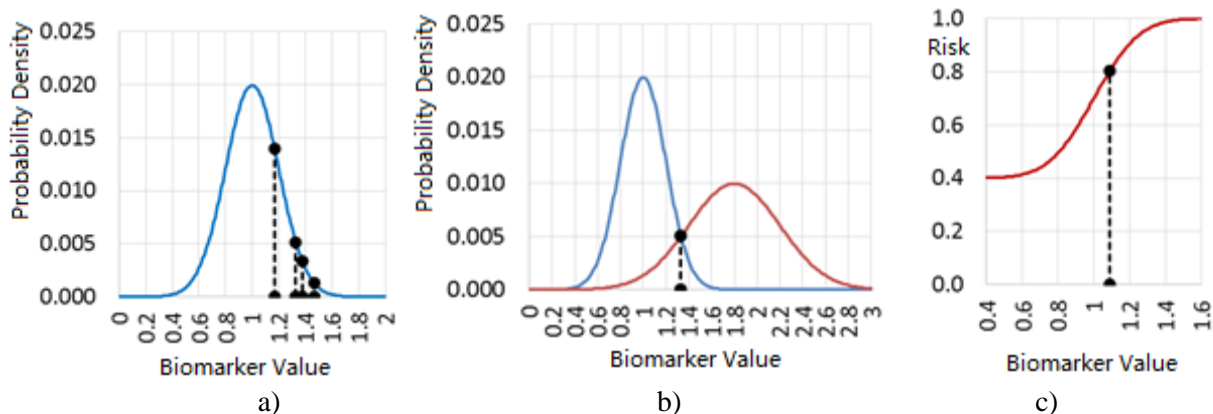

a)          b)          c)

**Picture 1 - Normal and Abnormal Biomarker Values a) acceptable levels of the confidence coefficient; b) discriminatory limits of normal and abnormal; c) risk threshold**

However, in practice, when monitoring indicators of an individual person or a narrow group of people, there occurs a task of tracking the indicators in a dynamic mode, as well as their approximation to or withdrawal from the rejection line. It can provide a prognosis of the indicator's value for some moment in the future and to prevent, possibly, its transfer to the critical range.

The area of Applied Statistics that considers mathematical aspects of biomarker data processing together with preparation of the conclusion variants is called Bioinformatics. It applies various methods and models, in particular:
- statistical estimation;
- pattern recognition;
- cluster analysis;
- multivariate regression analysis;
- sequential estimation (including the Bayesian one) etc.

When it is necessary to provide an accurate recognition of a person's state as well as to predict the coming changes, the important role is played by:

1. complex biomarkers, i.e. the interdependencies that provide a synergistic effect (for example, impact of a psychological state upon physiological indicators etc.);

2. consideration of the impact that additional factors (age, fitness, environment etc.) have upon the indicator's value.

Therefore, using the terminology of Applied Statistics, the interconnection of indicators and factors can be presented by different ways, in particular by:

- multivariate regression model (multiple regression);
- pattern recognition tasks;
- clustering tasks and others.

**Health Monitoring Data Analysis**

Fitness maintenance is one of the main tasks that defines working capacity of the team during longterm space expeditions. Fitness level is also defined by such an indicator as Heart Rate Recovery Time (HRRT) calculated after the exercises. HRRT of fit people is lower than of non-fit people. Heart Rate of fit people can vary between 50 to 240 heartbeats per minute depending on the exercise load. The machines that can assess a current value of HRRT are stationary bicycles, treadmills and other special machines where it is possible to set necessary test exercises and to measure current heart beat indicators, using which the HRRT is later estimated via standard or specialized software.

Modern equipment that is used to monitor physiological indicators often allows recording the measurement data on storage media, sending them via radio channels (Bluetooth, WiFi etc.) or in the Internet to a set email address.

Depending on the exercises and fitness level, the heart beat change graphs may look like Picture 2 when recovering.

For each of the graphs it is possible to compute a value of the only indicator that demonstrates the speed with which the heartbeat recovers. This indicator serves as one of the biomarkers of a person's fitness level. According to its values it is possible to recommend the physical exercises that would keep the person fit.

The heartbeat recovery process can be usually presented as an exponential relationship of the following type:

$$f(t) = a + (d - a)e^{-\theta t} + \varepsilon, \qquad (1)$$

where $a$ is a normal value of the heartbeat; $d$ is the heartbeat's value right after the exercises; $\theta$ is a speed coefficient of the heartbeat's recovery; $\varepsilon$ is a random component of the changes. According to current values of a heartbeat's target indicator $f(t)$ it is possible to compute the value of $\theta$ coefficient, using, for example, a least square method and specialized or standard software.

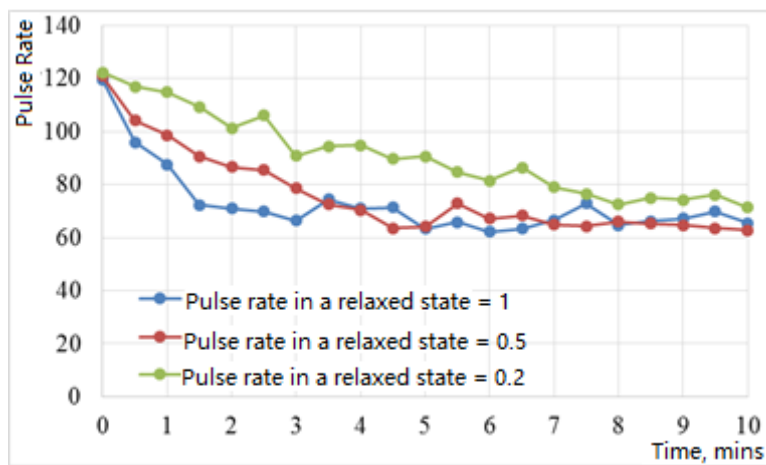

**Picture 2 - Heartbeat Recovery Process Measurement After Physical Exercises (for various fitness levels)**

Monitoring of $\theta$ coefficient can serve as the basis for correcting the fitness level of a person thus influencing his/her working capacity and environmental stability.

Other biomarkers, in particular frequency and depth of breathing are measured in the sport sphere, apart from the heartbeat. A list of possible biomarkers, whose values shall be measured in longterm space expeditions, can be rather long and expansive. Thus, current physical activity can be measured with microsized accelerometers (for example, three-axis micromechanical acceleration sensors) sewn into clothes in different places: on arms, legs, head etc. The sensors can transfer the data for its registration and processing via the radio channels. Statistical analysis of the data (for example: variable, correlation or a cluster one) can demonstrate which parts of the body receive a spontaneous (taken in the process of a current activity) or a special load.

Within the process of a continuous monitoring, such biomarkers (factors, input variables) should be connected with some target indicators (output variables) that depend on them or that are significantly correlated with them. Target indicators demonstrate level of a working capacity, error probability, power of concentration, disease resistance (immunity level) etc.

Within the process of the current data monitoring we can build dependencies of the indicators ($L$) from the factor vector ($\bar{x} = [x_1, x_2, \cdots, x_n]^T$) in the form of linear or non-linear regressional relationships $L = f(\bar{b}, \bar{x})$, where $\bar{b}$ is a parameter vector.

If we build the relationship in a linear form and with standard variables, the large values of coefficients $b_i$ shall indicate that their corresponding factor (biomarker) $x_i$ is a predictor, i.e. predicts significantly the indicator's value. Thus, in this case, using values of the predictor we can take controlling (correcting) measures without knowing the value of the indicator.

**Conclusions**

Taking into consideration the analysis presented in the work, it is possible, for example, to apply the technology of monitoring, analysis and health management of a group of people in a certain professional sphere, in particular in longterm space expeditions.

1. In order to build effective methods that would predict the state of a person who performs his/her specific professional duties, it is necessary to collect (using a real control group) an adequate amount of statistical data that would demonstrate a norm or a pathology (abnormality).

2. While operating at a real data field of a subject domain under consideration, it is necessary to discover the most significant biomarkers, predictors and their most effective complexes.

3. In order to provide a maximum accuracy of assessment and forecasting, to examine the dependencies of significant biomarkers and/or predictors (including the complex ones) from the factors of external and internal environment that have a significant impact upon their prognostic qualities.

4. To examine and determine characteristics of the assessment accuracy and prediction of the indicators that demonstrate changes in the state of study group members.

5. To examine the degree of a possible personification of models and methodology of the specific subject's health state analysis.